\documentclass{article}

\usepackage{chicago}

\newcommand{\iec}{\mbox{i.\,e.\,}}

\newcommand{\egc}{\mbox{e.\,g.\,}}

\newcommand{\dr}[1]{\ensuremath{\mathrm{d} #1\,}}

\newcommand{\ket}[1]{\ensuremath{\left|  #1 \right\rangle}}
\newcommand{\bra}[1]{\ensuremath{\left\langle #1 \right|}}

\newcommand{\proj}[2]{\ensuremath{\ket{#1} \bra{#2}}}

\newcommand{\matel}[3]{\ensuremath{\bra{#1} #2 \ket{#3}}}
 
\newcommand{\op}[1]{\ensuremath{\widehat{\textsf{\ensuremath{#1}}}}}

\newcommand{\denop}{\ensuremath{\rho}}

\newcommand{\tr}{\textsf{Tr}}

\newcommand{\be}{\begin{equation}}
\newcommand{\ee}{\end{equation}}

\begin{document}

\title{Inferential vs. Dynamical Conceptions of Physics}
\author{David Wallace}
\maketitle

\begin{abstract}
I contrast two 
possible attitudes towards a given branch of physics: as \emph{inferential} (\iec, as concerned with an agent's ability to make predictions given finite information), and as \emph{dynamical} (\iec, as concerned with the dynamical equations governing particular degrees of freedom). I contrast these attitudes in classical statistical mechanics, in quantum mechanics, and in quantum statistical mechanics; in this last case, I argue that the quantum-mechanical and statistical-mechanical aspects of the question become inseparable. Along the way various foundational issues in statistical and quantum physics are (hopefully!) illuminated.
\end{abstract}

\section{Introduction}

Sometimes progress, especially in foundational matters, can come not from resolving a dispute but by clarifying its structure. This paper attempts to do this in the cases of foundations of statistical mechanics (SM) and quantum mechanics (QM). Its main theme is that we can identify two attitudes to a given area of physics --- \emph{inferentialism}, the idea that a given theory is a tool to allow us to make inferences about present and future facts or experiments, and \emph{dynamicism}, the idea that a theory is an account of the dynamical behaviour of systems entirely independent of our our own knowledge --- and that the divide between inferentialism and dynamicism illuminates the debate both in statistical and in quantum foundations.

Partly this hoped-for illumination occurs within each separate subject: I will try to show that the inferential-vs-dynamical dispute naturally captures much of the basic disagreement within statistical mechanics, and that discussions of the quantum measurement problem often presuppose one or other conception even in stating the \emph{problem}, so that proposed \emph{solutions} can be misunderstood. More interestingly, it illustrates the possibility of close links between strategies in one case and in the other.

I make the case for this in three parts. In section~\ref{SM} I explain how the two rival conceptions play out in classical statistical mechanics, and canvass the main problems with each. In section~\ref{QM} I do likewise for quantum mechanics, and consider by analogy how those ``main problems'' look in a quantum-mechanical context. At this point, I hope to have established strong \emph{similarities} between the inferential/dynamical dispute in the two fields.

However, it is possible to go beyond similarities. When we consider not classical but quantum statistical mechanics, there is  an almost complete collapse of the statistical-mechanical questions onto the quantum questions, to the point that it becomes essentially impossible to adopt the inferential conception in the one case and the dynamical conception in the other. In the process of establishing this in section~\ref{QSM}, I will argue that while classical statistical mechanics can be considered as a probabilistic extension of classical mechanics, the same is not at all true in quantum statistical mechanics.\footnote{I elaborate on this point in \citeN{wallaceprobstat}.}

I conclude that a single interpretative question --- whether to conceive of a given field in physics as a form of inference or as a study of dynamics --- plays a central role in the foundations of quantum theory, and \emph{the exact same role} in the foundations of statistical mechanics once it is understood quantum-mechanically. I conclude by drawing some morals for the study of the conceptual foundations of both fields.

\section{Conceptions of Statistical Mechanics}\label{SM}

In trying to get clear on what (classical) SM is, we can identify two main themes. The first starts from the observation that in macroscopically large systems we are necessarily ignorant of many features of the system:
\begin{enumerate}
\item We do not know its present microstate.
\item We do not know its exact Hamiltonian.
\item Even if we did know (1) and (2), we do not know how to solve the exact equations of motion in order to predict the system's future microstate.
\end{enumerate}
SM, on this account, is concerned with how we can make correct, or at any rate reasonable, inferences about macroscopic systems in the face of these epistemic limitations. SM, that is, is a branch of the general theory of \emph{inference} under conditions of imperfect information. To quote a classic account:
\begin{quote}
The science of SM has the special function of providing reasonable methods for treating the behaviour of mechanical systems under circumstances such that our knowledge of the condition of the system is less than the maximal knowledge which would be theoretically possible. The principles of ordinary mechanics may be regarded as allowing us to make precise predictions as to the future state of a mechanical system from a precise knowledge of its initial state. On the other hand, the principles of SM are to be regarded as permitting us to make reasonable predictions as to the future condition of a system, which may be expected to hold on the average, starting from an incomplete knowledge of its initial state.\footnote{See footnote \ref{tolmanfn} for source.}
\end{quote}
The other concept begins from the empirical observation (going back at least to the nineteenth-century development of thermodynamics) that the collective degrees of freedom of macroscopic systems demonstrate observed, lawlike regularities: ice cubes melt in water, gases expand to fill boxes, heat flows along metal bars in accordance with the diffusion equation, and so forth. On this conception of SM, its goal is to derive these collective dynamical results from the underlying microdynamics. In principle this might be an exceptionless derivation, much as we can derive exceptionless equations for the movement of the centre of mass of a general body, or the rotational dynamics of a rigid body, even when that body is made of a very large number of components. But in practice there are strong reasons to expect any such statistical-mechanical derivation to require additional assumptions to be made. 

On this account, SM is a branch of the general theory of \emph{dynamics}: it studies a sub-problem of the general problem of understanding the behaviour of systems over time. A classical statement of \emph{this} position is the following:
\begin{quote}
[I]n the case of a gas, consisting, say, of a large number of simple classical particles, even if we were given at some initial time the positions and velocities of all the particles so that we could foresee the collisions that were about to take place, it is evident that we should be quickly lost in the complexities of our computations if we tried to follow the results of such collisions through any extended length of time. Nevertheless, a system such as a gas composed of many molecules is actually found to exhibit perfectly definite regularities in its behaviour, which we feel must be ultimately traceable to the laws of mechanics even though the detailed application of these laws defies our powers.For the treatment of such regularities in the behaviour of complicated systems of many degrees of freedom, the methods of statistical nechanics are adequate and especially appropriate.
\end{quote}
The two conceptions are not always cleanly separated --- indeed, my two quotes come from consecutive pages of the same classic textbook\footnote{\label{tolmanfn}\citeN[pp.1-2]{tolman}} --- and I don't necessarily want to claim they are incompatible, but they can have very different foundational implications, as a few examples should make clear.

\subsection*{The conceptual status of probability}

On the inferential conception, we are assumed to have imperfect knowledge of at least the current microstate. It is therefore natural mathematically to introduce probability measures to represent this imperfect knowledge: the probability assigned to a given microstate represents our level of confidence that the system is really in that microstate. To borrow a very useful piece of terminology from the quantum foundations literature: we can distinguish the \emph{ontic state} of the system, which represents what properties and features the system actually has, from the \emph{epistemic state}, which represents our ignorance of the ontic state. The ontic state is given by a phase-space \emph{point}, the epistemic state by a phase-space \emph{probability distribution}. (There is then a secondary question about how constrained our choice of epistemic state is: on the influential `objective' approach championed by Edward Jaynes\footnote{See, \egc, Jaynes~\citeyear{jaynesstatmech,jaynesstatmech2} and the papers in \citeN{jaynes1983}.}, it is specified uniquely, for given 
information, by a combination of maximum-entropy principles and considerations of symmetry; on a more `subjective' approach, two agents with the same information might rationally disagree. I will be concerned here largely with Jaynes' version of the inferential conception, which has been dominant in the physics literature for some years.)

On the dynamical conception, there is no up-front need for probabilities. We are interested in the large-scale features of the system, and this may require \emph{statistical} considerations --- what fraction of molecules in a gas have a given velocity, for instance --- but these can be understood as categorical features of the system. Boltzmann's original discussion of the H theorem is generally thought to have had this form, in particular (see \citeN{brownboltzmann} and references therein for historical discussion): Boltzmann took himself to be deriving, under reasonable assumptions, the conclusion that the statistical distribution of molecular velocities in a  dilute gas would \emph{reliably} approach  the Maxwell-Boltzmann distribution.

However, it is well known that the objections of Zermolo and Loschmidt established that Boltzmann's results could not be exceptionless facts about the dynamics of dilute gases, but could only hold given certain other assumptions --- and Boltzmann's, and later, attempts to fill in these assumptions have invariably turned out to be probabilistic, at least in part. That is, the dynamical conception ends up making claims not about how a system will \emph{invariably} behave but about how it will \emph{most probably} behave.  Furthermore, while \emph{this} requirement for probability is a requirement for a probabilistic microfoundation for a deterministic macroprediction (the approach to equilibrium of dilute gases), plenty of the applications of SM (notably, fluctuation phenomena) also lead to probabilistic predictions.\footnote{For more on this point see \citeN{wallaceactualstatmech}.} So while the dynamical approach does not require probability \emph{a priori}, it has in fact proven to be an essential component of SM. Indeed, the machinery of contemporary SM makes very extensive use of probability measures over phase space: any attempt to make sense of that machinery must make sense of those probabilities.

But on the dynamical conception, how are these probabilities to be understood? Not as representations of our ignorance of the true microstate (on pain of collapsing the dynamical conception into the inferential one). For some time a popular suggestion was that they should be understood as long-time averages, but this is beset with conceptual and technical problems, notably in the interpretation of non-equilibrium SM. The language of ``ensembles'' suggests that they are to be understood as relative frequencies in a fictional, infinite, collection of copies of the system --- but if the collection is fictional, how are we to understand its relation to the single, actual, system? If there is anything \emph{physical} in the world corresponding to these probabilities, the only obvious candidate is relative frequencies in actual systems --- but there are well-known problems\footnote{See, \egc, Hajek~\citeyear{hajek1996,hajek2009}.} in identifying probability with frequency in this way.

(This is perhaps a good point to note that the division between inferential and dynamical conceptions of SM is often portrayed in the literature as a division between Gibbsian and Boltzmannian versions of SM (see, for instance, \citeN{Callender2002} and \citeN{frigginrickles}.) This confuses the use of certain bits of mathematical machinery with the interpretation placed on that machinery; and unhelpfully suggests that a defender of the dynamical conception of SM has to reject use of probability distributions, and indeed that they have to reject or reconstruct a large fraction of contemporary results in SM. In \citeN{wallacegibbsboltzmann} I argue that Gibbsian and Boltzmannian mathematical methods can both be understood within a dynamical perspective, and indeed that from that perspective the differences between them are relatively minor.) 

\subsection*{Equilibrium}

The foundational assumption of thermodynamics, inherited by equilibrium SM, is that isolated systems can in general, and perhaps after a certain period of time has elapsed, be assumed to be in \emph{equilibrium}: to be in a state whose macroscopic parameters are unchanging with time. SM offers a concrete characterisation of this equilibrium in terms of a a probability distribution ---  the \emph{microcanonical ensemble}, in the simplest case ---  and it is a well-confirmed empirical fact that measurements of the values of macroscopic parameters of equilibrium systems, and of the fluctuations in those values, are accurately predicted by that distribution.\footnote{The \emph{Boltzmannian} characterization of equilibrium defines a system as at equilibrium not when its probability distribution is microcanonical but when its microstate is located in a certain phase-space region. But even Boltzmannians need to recover fluctuation phenomena, and so must require that an isolated system left to its own devices for the equilibration timescale has a probability of being in a given phase-space region given by the microcanonical distribution. At this point it seems to be largely a matter of semantics how we define equilibrium and whether we characterise fluctuations as occuring at equilibrium or into and out of equilibrium. (I make this case in more detail in \citeN{wallacegibbsboltzmann}.)} Both inferentialists and dynamicists must account for these facts.

They do so, however, in very different ways. For inferentialists, that a system is at equilibrium is a largely \emph{a priori} matter. According to the maximum-entropy principle defended by Jaynes (and finessing certain measure-theoretic concerns), the equilibrium probability distribution is the rationally required distribution that represents my ignorance of the microstate of a system when all I know of that system is its energy.  To say ``system X is at equilibrium'', for the (Jaynesian) inferentialist, is to say just ``all I know of system X is its energy''. As such, the question of how systems approach, or end up at, equilibrium, is thus not obviously well-posed on the inferentialist conception: whether a system is at equilibrium or not is a property not of the system alone but of the observer's knowledge of the system.

For dynamicists, on the other hand, whether or not a system is at equilibrium is a contingent claim about that system alone, and the claim that isolated systems should in general be treated as being at equilibrium is justified only if it can be established (on whatever additional assumptions are required) that isolated systems initially out of equilibrium approach equilibrium in a reasonable length of time (at least with high objective probability, however that is to be understood). The microcanonical characterisation of equilibrium is legitimate only if it is the unique probability distribution to which other probability distributions evolve in isolated systems (perhaps at some coarse-grained level of approximation rather than exactly). (Simlarly, the Boltzmannian claim that equilibrium is characterised by the largest macrostate is legitimate only if it is dynamically the case, on reasonable assumptions, that an isolated system's microstate is highly likely to evolve into that macrostate.) The foundations of equilibrium, on the dynamical conception of SM, thus require (in principle) quite detailed considerations of a system's dynamics: the Boltzmann equation can be seen as an early prototype of these kind of considerations, and the longrunning explorations of ergodicity and mixing  can be seen to be more modern attempts (however well or badly motivated those particular strategies might be).\footnote{For detailed consideration of such strategies, see \citeN[chapter 7]{sklarstatmech}. }

\subsection*{Thermodynamic entropy and the Second Law}

On the inferential conception, the Gibbs entropy,
\be 
S_G(\rho)=-k_B\int\dr{x} \rho(x) \ln \rho(x),
\ee
is normally identified with the thermodynamic entropy. This entropy, being a functional of the probability distribution, represents (on the inferential conception) a measure of an agent's lack of information about the underlying microstate: the more widespread and uniform the probability distribution, the higher the entropy. This interpretation of entropy as a measure of negative information can be formalised: on certain assumptions (notably including a designation of the Liouville measure as the preferred measure of uniformity on phase space) and up to a constant factor, it can be shown to be the unique such measure \cite{jaynesstatmech}.  It is maximised, under the constraint that the energy has a given fixed value, by the microcanonical distribution, and this provides the justification for the inferentialist of using that distribution to represent equilibrium.

The Gibbs entropy is well known to be invariant under Hamiltonian time evolution. It follows that if a system initially at equilibrium at energy $E$ and external parameters (say, volume) $V$ (that is, a system about which the observer knows only the energy and the volume) is allowed to evolve under Hamiltonian flow (perhaps involving external potentials) until it has energy $E'$ and volume $V'$, then the Gibbs entropy of the resultant distribution is less than or equal to the Gibbs entropy of the equilibrium distribution defined by $E'$ and $V'$. It has been argued \cite{jaynesstatmech} that this provides a justification for the 2nd Law of Thermodynamics on the inferential conception. The obvious problem with this account is that while the new \emph{equilibrium} distribution has higher entropy than the original \emph{equilibrium} distribution, the \emph{actual} new distribution --- which represents the observer's information about the current state given that at an earlier time it had energy $E$ and volume $V$ --- has the same entropy as the original distribution. The increase in entropy seems to occur because the observer discards some information (that the system came to have energy $E'$ and volume $V'$ in a specific way), rather than because of any feature of the system itself. Possible solutions to this problem include appeal to our lack of knowledge of the exact dynamics \cite[pp.353-4]{peres} or our lack of ability to perform the actual calculations that generate the time-evolved distribution or even to store the information about it \cite{myrvold2012}.

The Gibbs entropy does not seem suitable to represent thermodynamic entropy on the dynamical conception: there the probabilities must be understood (somehow) as objective features of the system, and so the Gibbs entropy is a constant of the motion. There are basically two resolutions of this problem. Firstly, we can define a coarse-graining map (or Zwanzig projection; cf \cite{zwanzig}) $J$ that smooths out the fine details of a probability distribution $\rho$, and define the \emph{coarse-grained Gibbs entropy}
\be 
S_{G;J}(\rho)=-k_B \int\dr{x} (J\rho)(x) \ln (J\rho)(x).
\ee
It is important to note that this is still a functional of $\rho$: $J$ is to be understood as a mathematical operation used in the definition of the entropy, not as a literal transformation of the underlying probabilities. (Note that on the \emph{inferential} conception this latter interpretation of $J$ might be acceptable --- representing, say, our finite powers of resolution of the system's detail --- but it is incompatible with the \emph{dynamical} conception as long as the underlying dynamics are taken to be Hamiltonian.)

The second resolution rejects the Gibbs entropy entirely, and instead adopts Boltzmann's old definition: phase space is divided into cells (`macrostates') and the Boltzmann entropy $S_B(M)$ of a macrostate is $k_B$ $\times$ the logarithm of its phase-space volume. The Boltzmann entropy $S_B(x)$ of a \emph{microstate} $x$ is then just the Boltzmann entropy of the unique macrostate in which $x$ lies: this definition of entropy is entirely non-probabilistic. (For this reason, it is often argued (\citeNP{alberttimechance,Callender2002,goldsteinboltzmann,lebowitz07}) that the Boltzmann definition is preferable, and indeed that the Gibbsian definition only makes sense on an inferentialist conception of SM; I criticise this view in \citeN{wallacegibbsboltzmann}.)

In both cases, there is a considerable technical project left to carry out. Firstly, the correct notion of coarse-graining or macrostate partitioning must be found. Often the criteria for this notion are stated in epistemic terms (states are in the same macrostate if they are ``macroscopically indistinguishable'' or somesuch), but this is an unnecessary concession to inferentialism: on the dynamicist conception, the criterion for a coarse-graining being correct (as in any case of emergence) is simply that we can write down robust dynamical equations for the collective degrees of freedom which abstract away from irrelevant microlevel details.

Secondly, we have to show, or at least make plausible, that entropy thus defined does indeed increase under the transformations typical of thermodynamics. Neither $S_{G;J}$ nor $S_B$ is a constant of the motion, so there is no \emph{a priori} barrier to showing that they increase in such circumstances, but to show that they in fact \emph{do} increase requires engagement with the dynamical details of the system (for all that it can be made extremely plausible in many cases).

The significantly larger technical burden of accounting for the Second Law on the dynamicist conception might be taken to be a strength or a weakness as compared to inferentialism: a weakness, if you regard it as undesirable to have to get tangled up in the messy details of the dynamics; a strength, if you believe that the validity of thermodynamics must ultimately lie in the dynamics and that attempts to bypass the messy details have ``the advantages of theft over honest toil''\cite[p.71]{russell1919}.

\subsection*{Retrodiction and time asymmetry}

The underlying microphysics, under either conception of (classical) SM, make no particular distinction between past and future; but the actual universe shows, and SM models, manifestly time-asymmetric processes. The clearest example here is the approach to equilibrium: systems not currently at equilibrium generally evolve to equilibrium (and we have seen how each conception of SM attempts to explain this); why, by parity of reasoning, should we not expect that systems currently not at equilibrium were at equilibrium in the past? More generally, insofar as SM underpins non-equilibrium dynamical processes like the expansion of gases into an empty space, why does that reasoning not likewise work into the past?

On the inferential conception, the paradox might be put as follows: let's stipulate that given our information about the present-day state of a system, it is reasonable to infer that we should have a high degree of belief that the system is in such-and-such state in the future. (For instance, suppose that our current information about a glass of water is that it contains an ice cube and some warm water (of given volumes and average energies); let us stipulate that it is reasonable on that information to infer that time $+t$  from now the ice cube is melted and the water is cooler. Why is it not equally reasonable to infer the same facts at time $-t$?

There is, on the face of it, a fairly straightforward answer. In fact, in realistic situations we have a great deal of information about the glass of water over and above its present state: we probably know what its state was five minutes ago, for instance, and even if we do not, we know a great deal of other facts about the state of the world five minutes ago. It turns out (and this is much of what gives SM its power) that this additional information about the past is of negligible importance in making predictions about the glass's future state --- but it is crucially important in making predictions about its past state. The asymmetry in our \emph{inferences} is caused by an asymmetry in our \emph{information}.

We can question whether we really have information about the past. After all, we have no \emph{direct} access to the past, but only to our memories of it, and those are presumably coded in the present state of the world (in particular, in our brains and our external records). And if all information is ultimately present-day information,  the `asymmetry of information' explanation for the asymmetry of inference seems to be in trouble.

Here's another way to put the same problem, which makes more stark its paradoxical aspects. There are good reasons to think that, conditional on the present-day macroscopic facts about the world and on the uniform (Liouville) probability measure over microstates consistent with that macrostate, that it is overwhelmingly more likely for our records of apparent past events to have spontaneously fluctuated into existence than it is for them to be consequences of those apparent events. So what justifies our belief that the past really happened?

Put this way, we seem to have a problem in epistemology rather than one in physics: the position that the past did not happen, after all, seems to be a form of scepticism, and the question of whether we are justified in treating our memories as information directly about the past or as information about our present-day brain state is a question about how to set up our epistemology. But something strange has happened here. For the facts that ice melts in water, that stars radiate light rather than absorbing it, that people grow old, and so forth, certainly appear to be facts about the world rather than about our means of inferring things about the world. So any explanation of them which relies for a success on a certain analysis of our epistemology seems problematic. Furthermore, as a matter of logic no such explanation can explain \emph{why} we as agents and as observers are in an asymmetric epistemic position. If observers, too, are just physical systems, we ought to explain why the system as a whole (including, inter alia, both the glass of water and the physicist contemplating it) displays the asymmetry in time that it does.

It's tempting at this point just to shrug and say that such an ineliminable role for the observer was always part and parcel of inferentialism. But this is too quick: the `inferential conception' is here a conception of \emph{SM}, not of physics as a whole. Recall the distinction between ontic and epistemic states: the real, objective, physics concerns the ontic state and its dynamics, and is represented by classical mechanics. SM is simply an inferential layer applied to that underlying reality to address our imperfect information about it. The problem for the inferentialist is not that their program makes use of notions like `observer' and `equilibrium': it is that various facts about the world --- and, in particular, its apparent time asymmetry, even in situations where no human intervention is occurring --- seem to belong more naturally to the observer-independent, ontic part of the theory, not to the epistemic part.

What of the dynamical conception? Here the paradox is more direct, because the question of whether we can derive a given time-asymmetric result is a question about objective facts about the world. If, for instance, at time $0$ we can derive on certain time-symmetric and time-translation-invariant assumptions that entropy $S(t)$ (either Boltzmannian or coarse-grained-Gibbsian) satisfies $S(+t)>S(0)$ for time $t>0$, the symmetry of the underlying dynamics tells us that $S(-t)>S(0)$ for $t>0$ also. And then by time \emph{translation} symmetry $S(0)>S(+t)$, and we have no paradox, but a straightforward algebraic contradiction. So our derivation must build in some violation of time-reversal or time-translation invariance and it is simply a matter of finding it.

This task is, in fact, straightforward. The dynamicist approach to probability requires an assumption that the probability distribution at the initial time has certain features (such as being uniform across a given set of microstates), and these features are not in general preserved under time evolution. If we impose this required assumption at a given time, it will deliver the required dynamical results at later times but not at earlier times. So insofar as the assumption is to be thought of as a \emph{physically contentful} boundary condition, and not simply a statement about our information, it must follow that we have to impose that boundary condition at the earliest relevant time for the system under study. And since boundary conditions cannot be set for each system separately in an interconnected universe, ultimately (as most advocates agree) the dynamical conception requires a particular condition to be imposed on the initial state of the Universe. (The most common choice of such a condition is the dual requirement that (i) the initial macrostate of the Universe was a particular low-Boltzmann-entropy state (the ``Past Hypothesis''), and (ii) the initial probability distribution is uniform or at least reasonably smooth over that macrostate; in \citeN{wallacelogic} I argue that (i) is not in fact needed.)\footnote{It is perhaps worth noting that defences of the Past Hypothesis often seem to equivocate between dynamical and inferential conceptions of SM: often the probability measure seems to be argued for \emph{a priori} and the Past Hypothesis is then justified on epistemic grounds, to avoid global scepticism.}

\subsection*{Compare and contrast?}

So: in the cases of probability, of equilibrium, of entropy, and of time asymmetry --- that is, in the main loci of discussion in foundations of SM --- the inferential and dynamical conceptions of SM give very different accounts of what is going on.  Which conception works best? So far as I can see, the most severe problems with each are as follows:
\begin{itemize}
\item The most serious problem for the dynamical conception is \emph{probability}. SM is suffused with probabilistic claims and couched in probabilistic language. Yet it is extremely difficult to see just how these probabilities are to be understood, if not as some quantification of our imperfect information about the actual state. And what role can a probability distribution over such states play in an explanation of the actual dynamical behaviour of the system, given that how it evolves in the future depends entirely on its actual, unique, state?

(Note that this is not simply the general philosophical problem of how objective chance can be understood in science. It is one thing to say of a system whose present state is $x$ that it has a certain objective probability of transitioning in the next instant to a state $x'$. It is quite another to say of that system that it has a certain objective probability of \emph{currently being} in state $x'$.)

Advocates of a dynamical conception are sensitive to this worry --- it drives the widespread scepticism about Gibbs entropy and the corresponding preference for Boltzmann entropy --- but simply defining entropy that way will not, by itself, suffice to remove the reliance on probability in SM.
\item The most serious problem for the inferential conception is \emph{objectivity}, and the most dramatic and serious example of the problem is asymmetry in time. To the inferentialist, virtually none of the claims made in SM are claims about the world in itself, but just about how I should reason about the world given imperfect information. This already has a somewhat problematic feel in the inferentialist \emph{characterisation} of equilibrium not as a state which systems generally speaking \emph{in fact} get into and remain in, but just as a way of saying that we don't know anything about the system's state. It comes close to paradox when we ask for an account of why the \emph{non-equilibrium} processes in the world display a clear and consistent time asymmetry, and in particular if we ask for an account of that time asymmetry that does not make essential reference to an external, already-assumed-asymmetric observer.
\end{itemize}

This should not be surprising. Classical SM seems to be a hybrid, displaying some features that suggest an inferential conception and some a dynamical one. Probability, in the classical deterministic context, is extremely difficult to understand dynamically. The time-asymmetric dynamics of the Boltzmann equation and its many relatives is extremely difficult to understand inferentially. (Equilibrium thermodynamics could be called either way: is it the study of which processes are \emph{physically possible}, or which transformations \emph{are within the power of an agent}?)

If classical mechanics were correct (and if, \emph{per impossibile}, I could still exist under that assumption), I would end here with the suggestion that the inferentialist-vs-dynamicist way of understanding the debates in SM is more helpful, and less prone to mutual miscommunication, than that Gibbs-vs-Boltzmann approach currently prevalent. However, the move from classical to \emph{quantum} SM radically changes the terms of the debate, as we will see. Firstly, though, it is necessary to consider quantum theory \emph{itself} from inferential and dynamical perspectives.

\section{Conceptions of Quantum Mechanics}\label{QM}

The basic dynamical axioms of QM are simple enough to state, and can be done in direct parallel with those of classical mechanics. Instead of a state space of phase-space points, we have Hilbert-space rays. Instead of evolution under Hamilton's equations, we have evolution under the Schr\"{o}dinger equation. And to find the state space of a composite system, instead of taking the Cartesian product of phase spaces we take the tensor product of Hilbert spaces.

There is, however, a different parallel we could have drawn. Quantum states could have been considered to be analogs of \emph{probability distributions} over phase space, not of phase space points; the Schr\"{o}dinger equation could have been compared to the Liouville equation, not Hamilton's equations; the tensor-product rule for constructing the state spaces of composite systems could have been regarded as correct in QM \emph{just as} in classical probabilistic mechanics.

At the root of the difference between the two approaches is this question: is the quantum state something like a physical state of a system, or something like a probability distribution? One way\footnote{A somewhat heterodox way, to be sure; I develop it more fully in \citeN{wallacemodern}.} to state the notorious \emph{measurement problem} is that ``orthodox'' QM --- that is, QM as it is in practice used\footnote{``Orthodox QM'' is often used instead in the foundational literature to refer to the Dirac-von Neumann concept \cite{Dirac1930,vonneumann} of two different quantum dynamics: unitary evolution under the Schr\"{o}dinger equation most of the time, stochastic collapse during measurement. On this approach, the state is unequivocally non-probabilistic in nature but the dynamics are not reliably unitary. The approach is generally described as orthodox in the sense that it represents ``textbook QM''. In reality, it very much depends on the textbook, but in any case (see again \citeN{wallacemodern}) it does not do a good job at representing current practice in the \emph{use} of QM.} --- systematically equivocates between the two ways of understanding the state:
\begin{description}
\item[The state as probability distribution:] The quantum state of macroscopic systems, and the quantum state of even microscopic systems in contexts of measurement, is treated as representing a probability distribution over possessed values: if a system is an a macroscopic superposition of, say, `cat alive' and `cat dead' states, we treat the cat as \emph{either} alive \emph{or dead}, with the probability each equal to the mod-squared amplitude of the corresponding state in the superposition. Similarly, we analyse contexts of state \emph{preparation} as something very akin to probabilistic conditionalisation: if we generate a collimated beam of atoms by putting a narrow slit in front of a furnace, we conditionalise on the fact that the particles went through the slit, and discard the part of the quantum state corresponding to the atoms \emph{not} going through the slit.\footnote{I am grateful to Simon Saunders (in conversation) for pressing the role of state preparation here.}
\item[The state as representing something physical:] The quantum state of microscopic systems, or indeed any systems, in situations where interference phenomena occur, is (prima facie) treated as representing a physical, albeit in general highly non-classical, state of the world. In an interference experiment, we talk freely about interference between the parts of the quantum state corresponding to the various terms in the superposition, in a way that does not straightforwardly lend itself to probabilistic reinterpretation.
\end{description}
In this way of looking at things, the ``measurement problem'' is simply this incoherence about how we are to understand the quantum state: what are the criteria for when we should treat it as a probability distribution and when as a physical state, and how can we make sense of the transition between the two conceptions?

In the space of moves made to resolve/solve/dissolve the measurement problem, I believe we can identify two broad themes, which roughly track the two themes discussed in \emph{statistical} mechanics; I discuss each in turn.

\subsection*{The inferential conception of QM}

On this conception, the quantum state does not represent anything physical: it is to be understood as a probability measure, and that probability measure represents in some way our degrees of belief in some set of physical facts. On this conception, macroscopic superpositions like  ``Schr\"{o}dinger-cat'' states are utterly unproblematic; I represent a cat with a Schr\"{o}dinger-cat state just in case I am unsure as to whether it alive or dead following some measurement process. Asher Peres expresses this position with great clarity:
\begin{quote}
[T]he ``cat paradox'' arises because of the naive assumption that the time evolution of the state vector $\psi$ represents a physical process which is actually happening in the real world. In fact, there is no evidence whatsoever that every physical system has at every instant a well defined state $\psi$ (or a density matrix $\rho$) and that the time dependence of $\psi(t)$ (or of $\rho(t)$) represent the actual evolution of a physical process. In a strict interpretation of quantum theory, these mathematical symbols represent different \emph{statistical information} enabling us to compute the \emph{probabilities} of occurrence of specific events.
\end{quote}
Peres also makes clear why I call this approach ``inferentialist'': it interprets QM not as a description of physical reality, but as a calculus to make (probabilistic) predictions in the absence of such a description. Chris Fuchs says it even more explicitly:
\begin{quote}
Quantum states are states of information, knowledge, belief, pragmatic gambling commitments,
\emph{not} states of nature.  \cite[section 6]{fuchsinformation}
\end{quote}  

If inferentialism dissolves the problem of macroscopic superpositions, problems arise when we apply it to \emph{microscopic} superpositions. We can in practice treat 
a Schr\"{o}dinger-cat state as a probabilistic mixture of live and dead cat without contradicting ourselves; if we try to treat an electron in a superposition of spin states as a probabilistic mixture of those states, we will get the wrong answers in our calculations. So if the probabilities are \emph{not} probabilities for the electron to have various values of spin, what \emph{are} they probabilities for?

By analogy with (inferentialist) SM, we can identify one apparently highly attractive possibility. In SM, we distinguished between the ontic state (representing the actual world) and the epistemic state (representing our information about, or degrees of belief about, the ontic state). It is tempting, then, to read the quantum state likewise as an epistemic state (indeed, what I call the inferentialist conception is often called the $\psi$-epistemic approach, as contrasted with the $\psi$-ontic approach which takes the quantum state $\psi$ as representing something physically real). 

This suggests a research program: find that theory which is to QM as classical microdynamics is to (inferentialist) SM. This is the program which Einstein, for instance, seemed to have in mind when he spoke of `hidden variables'.) But the progress made in that program, though substantial, has been largely\footnote{The only significant \emph{positive} result of which I am aware is Rob Spekkens' `toy model' \cite{spekkens}.} negative, notably:
\begin{itemize}
\item Bell's theorem requires that the theory involve instantaneous action at a distance.
\item The Kochen-Specker theorem requires that the theory have the apparently-pathological feature of \emph{non-contextuality}.
\item The recently-proved Pusey-Barrett-Rudolph theorem (\cite{puseyetal}; see \cite{Maroney2012} for discussion and development) rules out such theories entirely, given apparently-very-weak assumptions about our ability to treat systems as independent from one another.
\end{itemize}

The main alternative proposed is that the probability distribution is over possible outcomes of measurements, represented either by projection-valued measures on Hilbert Space (PVM), or (more commonly in recent work) by positive operator valued measures (POVMs). There is no mathematical obstacle to this move; indeed, it can be proven (in the case of PVMs, with great labour~\cite{gleason}; in the case of POVMs, fairly straightforwardly~(Caves \emph{et al}~\citeyearNP{cavesfuchsgleason}) that any such probability measure, provided that it is non-contextual,\footnote{Something admittedly difficult to justify on this conception of the state; cf \citeN[pp.225-6]{wallacebook}.} is represented by some pure or mixed quantum state.

The problem with \emph{this} strategy is that it is hard to understand it as anything other than straight instrumentalism. Some of its advocates (\egc \citeN{peres}) are happy to accept this; others (notably \citeN{fuchsinformation}) reject it and maintain that their proposal is compatible with some kind of realism, but it is obscure to me exactly what they hope to be realist about. (See \citeN{Timpson2008} for further discussion.)

\subsection*{The dynamical conception of QM}

On this conception, QM --- like classical mechanics --- is a dynamical theory, concerned with how the physical features of the world evolve over time without reference to any observer, and the quantum state is taken to be entirely objective. This more or less commits advocates to accepting (i) that the quantum state is not \emph{fundamentally} a probabilistic entity, and (ii) that at least in the case of microscopic systems, it needs to be understood as physically representing their properties in the same basic way as does the classical microstate.

The burning question\footnote{Philosophers have been almost as concerned with a separate problem: if the quantum state is a physical state, how are the physical features of that state to be understood? For discussion of this point, see, \egc, \citeN{maudlinconference}, \citeN{hawthorneconference} and \citeN{wallaceFAPP}, and the papers in \citeN{neyalbert}. (With rare exceptions, physicists have been fairly unconcerned.)} for this strategy is then: how come the quantum state \emph{appears} to be probabilistic when macroscopic systems are concerned? Here a large amount of technical progress has been achieved, under the general label of \emph{decoherence theory}. Two broad strategies have been applied here:
\begin{itemize}
\item The \emph{decoherent histories} (or consistent histories) program\footnote{See, \egc, Griffiths~\citeyear{griffiths,griffiths96,griffithsbook}, Omnes~\citeyear{omnes,omnes92,omnesbook}, Halliwell~\citeyear{halliwellhydrodynamic,halliwellconference}, \citeN{gellmannhartle93}, and \citeN{hartleconference}.} has asked directly: under what conditions do the probabilities assigned by QM to sequences of observables obey the probability calculus? The general conclusion is that at least a sufficient condition is that QM delivers probabilistic (or very-nearly-probabilistic) predictions for the evolution of coarse-grained properties of systems: that is, it is probabilistic on the macro scale once the microscopic details are averaged over.
\item The \emph{environment-induced decoherence} program\footnote{See, \egc, \citeN{zeh93},Joos \emph{et al}~\citeyear{joosetal}, Zurek~\citeyear{zurek91,zurek01review}, and \citeN{schlosshauerbook}.} has considered how the evolution of a macroscopic-scale system is affected by its coupling to an (internal or external) environment). The main focus of the program has been the selection by the environment of a preferred basis (usually a wavepacket basis) for the system with respect to which interference is strongly suppressed. This in turn has the consequence that the system can be consistently treated as probabilistically evolving. (This point is stressed by \citeN{zurekroughguide}; see also \citeN[ch.3]{wallacebook} for discussion.)
\end{itemize}
As a result, we now have a reasonably good understanding of the dynamical processes by which the quantum state comes to have the structure of a probability measure with respect to macroscopic degrees of freedom, even while being nonprobabilistic at the micro-level. (In the process, we also learn why the in-principle-incoherent shifting between probabilistic and non-probabilistic readings of the quantum state does not lead to practical problems.\footnote{See \citeN{wallacemodern} for further discussion of this point.})

It is at best controversial whether all this provides a \emph{conceptually} satisfactory understanding of probability in QM, and of QM more generally. If the quantum state ultimately is taken to represent something physical, then the terms in a superposition each seem to represent physical features of the world. In a Schr\"{o}dinger-cat state, then, the fact that the amplitudes of the live-cat and dead-cat terms can be consistently treated as probabilities does not seem to conceptually justify assuming that only one represents anything physically real. That is, the dynamical conception of QM --- at least as long as QM itself is unmodified at the formal level --- is tantamount to the Everett interpretation. In this context, the live-cat and dead-cat terms represent physically coexistent goings-on --- `branches' in the usual terminology --- and it has been extensively discussed whether more than decoherence is needed to justify interpreting the branches' weights probabilistically, and if so, whether and how the gap can be filled. (For an introduction, see \citeN{greavescompass}; for extensive discussion, see Saunders \emph{et al}~\citeyear{saundersetal}; I give my own account in \citeN[ch.4-6]{wallacebook}.)

The main alternative to the Everett interpretation within the dynamical conception is to modify quantum theory --- either by adding a dynamical state-vector collapse rule to eliminate all but one term in a macroscopic superposition, or by adding hidden variables to pick out one term as somehow preferred. (Notice that unlike the (ill-fated) inferentialist hidden-variable strategy, this version maintains the quantum state as physically real and adds additional hidden variables.) In dynamical collapse theories (of which the best-known are the Ghirardi-Rimini-Weber theory~\cite{grw} and the Continuous State Localisation theory~\cite{pearle}), probabilities now become part of the laws of physics, via an irreducibly stochastic dynamics. In hidden-variable theories (of which the de Broglie-Bohm theory is much the best known), the probabilities normally\footnote{At least, they do so in the case of the de Broglie-Bohm theory; as discussed in \citeN{bubbook}, it is perfectly possible to construct stochastic hidden variable theories where the probabilities have basically the same status as in dynamical-collapse theories.} enter via a measure over hidden variables whose interpretation recapitulates the puzzles of probability in classical SM. In either case the theory must be constructed so that the newly-added probabilities are numerically equal to the branch weights given at the coarse-grained level by the quantum state.

Philosophers have generally been more concerned than physicists about the conceptual difficulties of probability in the Everett interpretation, and less concerned about the technical difficulties incurred (especially in the relativistic case) by modifications of quantum theory. But both Everett and the modificatory strategy are examples of the dynamical conception: on both, the quantum state is resolutely physical; on both, physics is concerned with the actual dynamical behaviour of the world, independent of our knowledge of it.\footnote{This is perhaps somewhat controversial for deterministic hidden-variable theories, where there is space for an inferential/dynamical dispute about the nature of the probability measure over hidden variables.}

\subsection*{Compare and contrast? - the quantum case}

In my discussion of SM, I identified the most serious conceptual problems for the inferential and dynamical strategies as being, respectively, concerns about the objectivity of statistical mechanics' deliverances (in particular, the direction of time) and about the nature of statistical-mechanical probabilities if they are not to be understood epistemically. 
How do their quantum analogs fare?

It seems to me hard to deny that objectivity is a worse problem for quantum-mechanical than for statistical-mechanical inferentialists. In the statistical-mechanical case it was clear what part of physics was objective (classical mechanics) and what part was a matter of inference about the objective part (SM); the worry was that certain features of SM seemed to belong more naturally in the objective part. But it is extremely difficult, in the light of the various no-go theorems, to see what stands to classical mechanics as QM stands to SM. 

Put another way, QM is our current general dynamical framework, replacing classical mechanics (save in the gravitational regime). If that framework as a whole is to be understood inferentially, physics \emph{as a whole} seems to be an inferential framework, and it is no longer clear what we are inferring \emph{about}. 

As for the dynamical conception, probability is at least a very \emph{different} problem for  QM than for classical SM. In the latter case, the problem was that ``probabilities'' were just an additional layer placed over an actual, deterministic, underlying dynamics, and that as such it was very hard to understand them as actually representing an objective property of the physical system. In unmodified QM, probabilities are features of the quantum state in certain decoherent regimes, and there is no additional ``underlying dynamics'' beyond the dynamics of the quantum state. In the case of dynamical collapse theories, or of hidden-variable theories with stochastic hidden-variable dynamics, probabilities are instead the result of genuinely stochastic  laws. It is, of course, entirely possible to worry about the conceptual status of either explication of probability --- maybe we just can't make sense of probabilities in an Everett-type theory? maybe stochastic laws don't make sense? --- but it is at least clear what the analysis of probability would have to deliver, and clear that probability (whatever else it might be) is not an epiphenomenal gloss on underlying physics. (There is, in other words, no way of separating the theory into probabilistic and non-probabilistic parts.) Only in the special case of deterministic
hidden-variable theories do we see the problems of classical-statistical-mechanical probability reproduced in a basically-unchanged form.

\section{Quantum SM}\label{QSM}

So far I have drawn parallels between quantum and statistical mechanics, but the `statistical mechanics' I have considered is \emph{classical} SM. Since the world is quantum rather than classical, however, presumably the correct understanding of SM should proceed via quantum SM.

It is superficially tempting to suppose that we can map classical to quantum SM via a straight translation scheme: replace ``phase space'' by ``(projective) Hilbert space'', ``phase-space point'' by ``Hilbert-space ray'', and ``probability distribution over phase space'' with ``probability distribution over projective Hilbert space'', and then just apply the ideas of classical SM \emph{mutatis mutandis}. 
But I will argue that this is indefensible on a number of grounds.

To begin with a purely conceptual objection: classical microstates are in no way probabilistic, but on either the inferential or the dynamical conception of quantum physics, there is a probabilistic aspect to the quantum state, so that a probability distribution over quantum states is in a sense a probability distribution over probability distributions. This is particularly stark if we take an inferentialist approach to QM, and regard the quantum state as in some sense epistemic. In this case, only an inferentialist approach to \emph{statistical} mechanics seems to make sense, but then the probability distribution of SM is an epistemically-interpreted probability distribution over epistemically-interpreted probability distributions, and we might as well cut out the intermediate step. That is, if we take an inferentialist attitude to both quantum and statistical mechanics, the subject matter of the two disciplines is the same: both are concerned with our epistemic state. (We will shortly see how this plays out technically.)

Matters are somewhat more satisfactory \emph{conceptually} if we take the dynamical conception of QM, and regard the quantum state as representing the physical world. We now have a choice of interpreting the probability distribution \emph{over} quantum states as either an objective feature of those systems, or as epistemic. The former is at least inelegant: it requires two \emph{separate} conceptions of objective probability --- one to be understood via QM, the other via some unknown process. The latter is simplest to understand conceptually --- we treat QM, like classical mechanics, as the underlying dynamical theory, and regard statistical-mechanical probabilities as quantifying our ignorance of the dynamical state in each case. (For what it is worth, I have the impression that most authors commenting on the foundations of quantum SM think of it in this way.)

However, even if the two notions of probability in play are \emph{conceptually} distinct, they inevitably merge in any attempt to extract \emph{empirical} content from the theory. Suppose we assign a probability measure $\Pr(\psi)$ over quantum states $\psi$. Then the expectation value of any observable $\op{X}$ is given by
\be 
\langle \op{X} \rangle = \int \dr{\psi} \Pr(\psi) \matel{\psi}{X}{\psi}.
\ee
As is well known, if we define the \emph{mixed} state $\denop[\Pr]$ by
\be 
\denop[\Pr]=\int \dr{\psi}\Pr(\psi)\proj{\psi}{\psi},
\ee
this expression can be rewritten as 
\be 
\langle \op{X} \rangle = \tr(\denop[\Pr]\op{X}).
\ee
So all empirical predictions about the system are determined directly by the mixed state, and only indirectly by the probability distribution. This matters because the relation between probability distributions and mixed states is many-to-one. Even if there are two notions of probability present (one emergent from the quantum dynamics, one entering through SM) they are mixed together in a way which defies empirical separation.

Things become more dramatic still when we recall that so far we are working with a very impoverished notion of quantum state. As is well known, quantum systems can become entangled with their surroundings, and in doing so there is \emph{no} Hilbert-space ray that correctly represents the system. Furthermore, the systems typically studied by SM are macroscopically large, so that there is absolutely no reason to expect that such systems, even if initially prepared in pure states, will not rapidly become entangled with their environment. That is, we have excellent dynamical grounds to expect that the probability of the system initially being in \emph{any} pure state, even if it is initially taken to be 1, will rapidly approach zero.

There is a straightforward solution, of course: we can represent entangled systems by \emph{mixed} states (by taking the partial trace over the environment of the state of the combined system). So if we want to hold on to the idea of placing a probability measure over quantum states, that measure had best be defined over the space of mixed states (in which pure states are present only as a special case).

If such a measure  $\Pr(\sigma)$ is defined over mixed states $\sigma$, though, the many-to-one issue recurs in even sharper form. For all empirical predictions of the system are now given by the mixed state
\be 
\denop[\Pr]=\int \dr{\sigma} \Pr(\sigma) \sigma.
\ee
Any empirical result achieved by assigning to the system a \emph{probability distribution} over mixed states can be reproduced by assigning it a \emph{single} mixed state.

The introduction of probability measures over quantum states is thus epiphenomenal: any empirical prediction obtained via such a measure can be equally well obtained by assigning a single mixed state to the system --- and on the dynamical conception of QM (and given entanglement), mixed states are perfectly legitimate choices of state for a physical system. 

(Couldn't we hold on to the idea that physical systems only have pure states by rejecting the idea that entangled systems have states at all? But there is no reason to think that even the \emph{entire observable Universe} has a pure state (indeed, there are rather good reasons to think that it does not, since the Universe as a whole has a causal horizon and in quantum field theory these horizons generally give rise to thermal radiation, the state of which is mixed).

So: on the dynamical conception of \emph{quantum} mechanics, there is simply no need to introduce additional probabilities via \emph{statistical} mechanics, whether those probabilities are to be understood epistemically or in some more objective sense. The probabilities of quantum theory itself will do just fine. 

We have already seen that the same ought to be true, on conceptual grounds, if we adopt the inferential conception of QM. We can now see technically how this goes through: on the inferential conception there is even less reason to deny that a mixed state is a legitimate state of a system (indeed, Gleason's theorem might be interpreted as telling us exactly why it is the most general such state). A probability distribution over mixed states is then just a probability distribution over probability distributions, and should collapse to a single mixed state, as indeed it does.

The conclusion seems to go through on either conception. While \emph{classically} it might have seemed that classical SM is a probabilistic \emph{generalisation} of classical mechanics proper, quantum-mechanically it is just a \emph{restriction} of quantum mechanics to the regime in which statistical-mechanical methods (irreversibility, equilibrium, and the like) apply. \emph{Which} regime that is depends on whether the dynamical or inferential conception is adopted. On the former, it is the regime in which the methods used to derive irreversible \emph{dynamics} are applicable, and so is characterised \emph{inter alia}  by a large number of degrees of freedom and by an initial boundary condition that breaks the time symmetry.\footnote{This holds for most unitary versions of quantum theory, at any rate; in stochastic modifications of quantum mechanics, the time asymmetry might arise from the time-asymmetric stochastic dynamics. See \citeN{alberttimechance} and \citeN{wallaceinwilson} for further thoughts along these lines.}) On the latter, I have no idea, unless it is essentially characterised by external human intervention.

As a corollary, the questions of whether to adopt an inferential or a dynamical conception of physics cannot be answered independently in quantum and in statistical mechanics. The answer to the former determines the answer to the latter. And since classical mechanics is valid in our Universe only insofar as it is a valid approximation to quantum mechanics, these conclusions continue to hold true even in so-called ``classical'' statistical mechanics, which should be understood as quantum statistical mechanics in the classical limit. In particular, the probability distributions of classical statistical mechanics are quantum-mechanical states (pure or mixed) in a certain limiting regime. (On this point, cf Ballentine's observation \cite{ballentine} that the classical limit of quantum mechanics is classical \emph{statistical} mechanics; cf also \citeN{emersonequilibrium}, and my own observations in \citeN{wallaceprobstat}.)

As a case study of how QM and quantum statistical mechanics have essentially the same subject matter, consider again the direction of time. Both in non-equilibrium statistical mechanics (in, \egc, the Boltzmann equation) and in decoherence theory, physicists are in the business of deriving time-asymmetric dynamical equations from a time-symmetric starting point. (In both cases it is less than clear how to understand this from an inferential perspective.) We might expect, then, that there is no sharp divide between the two sorts of derivation. Indeed this turns out to be the case, as any perusal of the technical results in the respective fields illustrates. For instance, the decoherence master equation (a standard workhorse of environment-induced decoherence; see, \egc, \citeN{schlosshauerbook}) is derived by standard methods used in statistical physics to study dissipation; indeed, transformed to a phase-space representation the equation can be identified as simply the Fokker-Planck equation, a standard equation in kinetic theory (see, \egc, \citeN[p.301]{liboff}).

\section{Conclusion}

I have attempted to show that 
\begin{itemize}
\item[(i)] we can attempt to understand classical statistical mechanics  on either the inferential or the dynamical conception, where in either case the underlying classical microstate mechanics is understood dynamically; 
\item[(ii)] we can likewise attempt to solve the measurement problem in quantum mechanics according to either conception, although the (apparent) absence of any underlying dynamical theory separate from quantum theory significantly alters the debate; 
\item[(iii)] since quantum statistical mechanics can be understood as studying, directly, the quantum-mechanical states of individual systems (understood either inferentially or dynamically), the decision as to whether to understand quantum theory inferentially or dynamically forces the issue as regards the correct understanding of statistical mechanics.
\end{itemize}

I conclude by gathering together some general thoughts as to the implications for foundational and philosophical work in these areas:
\begin{itemize}
\item The current debate in the foundations of statistical mechanics, often characterised as ``Gibbs vs Boltzmann'', would be better characterised as ``inferential vs dynamical'', as the criticisms made of the Gibbsian approach by Boltzmannians are really criticisms of the inferential conception rather than of the Gibbsian \emph{machinery}.
\item The quantum measurement problem is helpfully understood as the problem of resolving a conceptual incoherence between probabilistic and non-probabilistic ways of understanding the quantum state, such that inferential (or $\psi$-epistemic) and dynamical (or $\psi$-ontic) approaches are different ways of resolving the incoherence: the former by treating the quantum state (somehow) as inherently representing an agent's probability function, the latter by treating probabilities as (somehow) emergent from a non-probabilistic underlying reality. Traditional ways of phrasing the measurement problem beg the question in favour of the dynamical conception and hinder communication. (This can still be the case even if the dynamical conception is \emph{right}.)
\item Although advocates of a $\psi$-epistemic view of the quantum state often advance their view by analogy to the classical probability distributions of statistical mechanics, this presupposes a controversial interpretation of classical statistical mechanics which in the quantum case collapses to a straightforward \emph{restatement} of the $\psi$-epistemic view.
\item While it may be of historical interest to understand how probabilities could have been understood in classical statistical mechanics considered in isolation, there is no point in seeking such understanding if our goal is to understand statistical mechanics in the actual world. In our world, the probabilities of statistical mechanics are just special cases of the probabilities of \emph{quantum} mechanics.
\item The only exception (I can see) to the above concerns the de Broglie-Bohm theory. Here it is commonly claimed that probability is to be understood just as in classical statistical mechanics. Advocates of the theory thus have a clear need (and thus, a clear motivation) to explore the interpretation and justification of probability in this context; they should not, however, be reassured by the supposed ``fact'' that ultimately they can piggy-back on the classical-statistical-mechanical explanation for such probabilities, as for all we know there may not be one.
\item Although it is a common strategy (frequently adopted, less frequently defended) to study the foundations of \emph{classical} statistical mechanics on the expectation that essentially the same issues arise in quantum statistical mechanics, this strategy is largely unfounded.
\item In particular, the idea that quantum statistical mechanics involves putting probability distributions over quantum states (just as classical statistical mechanics involves putting probability distributions over classical states) has no justification.
\item I will conclude on a conciliatory note. Although it is probably apparent that I am much more sympathetic to the dynamical than to the inferential conception of quantum mechanics, and of statistical mechanics, \emph{in general}, this need not mean that some \emph{areas} of these fields are much better understood on something more like the inferential conception. In quantum theory, it is fairly clear that information theory ought to be understood that way: its subject matter is not the dynamical behaviour of unattended systems, but the limitations imposed by physics on agents' activities. And in statistical mechanics, though the conceptual problems of statistical mechanics are often simply taken to be providing a microphysical foundation for thermodynamics,\footnote{See \citeN{wallaceactualstatmech} for criticism of this viewpoint.} ``thermodynamics'' is a misnomer: it is again concerned with agents' ability to control and transform systems, not primarily with how those systems behave if left to themselves.
\end{itemize}

\end{document}